\documentclass[12pt]{iopart}
\usepackage{epsf,epsfig,amssymb,graphicx,times}
\begin{document}
\title{Squeezed-state quantum key distribution upon imperfect reconciliation}
\author{Vladyslav C Usenko$^{1,2}$ and Radim Filip$^1$}
\address{$^1$ Department of Optics, Palack\' y University, 17. listopadu 12,  771~46 Olomouc, Czech Republic}
\address{$^2$ Bogolyubov Institute for Theoretical Physics of National Academy of Sciences,
Metrolohichna st. 14-b, 03680, Kiev, Ukraine}
\eads{\mailto{usenko@optics.upol.cz}, \mailto{filip@optics.upol.cz}}
\date{\today}
\begin{abstract}
We address the security of continuous-variable quantum key distribution with squeezed states upon realistic conditions of noisy and lossy environment and limited reconciliation efficiency. Considering the generalized preparation scheme and clearly distinguishing between classical and quantum resources, we investigate the effect of finite squeezing on the tolerance of the protocol to untrusted channel noise. For a long-distance strongly attenuating channel and the consequent low reconciliation efficiency, we show that feasible limited squeezing is surprisingly sufficient to provide the security of Gaussian quantum key distribution in the presence of untrusted noise. We explain the effect by behaviour of the Holevo quantity, which describes the information leakage, and is effectively minimized by the squeezed states. 
\end{abstract}
\maketitle

\section{Introduction}

Quantum key distribution (QKD) \cite{qkd1,qkd2} is well known to be the main practical application of the fundamental principles of quantum physics on the single quanta level and is the core part of quantum information processing. It has its goal in establishing communication channels, where physical principles guarantee the security of cryptographic keys. First proposed for the single quanta (qubits), it was lately developed on the basis of continuous variables (CV) \cite{cvqkd}, which are observables of either coherent \cite{coh2,coh3,coh4,coh5} or squeezed \cite{sq1,sq2,sq3,sq4,sq5} states of light. Due to use of quadrature modulation and homodyne measurement the latter approach does not need single-photon sources and detectors. In a typical CV QKD scenario, one of the trusted parties (Alice) performs displacement of a carrier state by modulating it in one or both of complementary quadratures. States are sent thought the channel, that is assumed to be fully under the control of a potential eavesdropper (Eve) and are then detected by the remote trusted party (Bob), which randomly measures one of the quadratures using a homodyne detector (or simultaneously both quadratures using a pair of the homodyne detectors in the balanced heterodyne scenario \cite{nonswitching,nonswitching2}). After transmitting a sufficient number of bits, parties exchange classical information and apply classical algorithms to perform reconciliation, which includes discretization of Gaussian data and error correction and results in identical bit strings, from which the secure key is distilled using classical privacy amplification algorithms. 

The unconditional security of the CV QKD protocols was proven for the general case of collective attacks \cite{coherent1,coherent2} (which implies security against the most general coherent attacks \cite{generalattacks}) using extremality of Gaussian states \cite{extremality} as such that provide maximum information to an eavesdropper. The lower bound on the unconditionally secure key rate in a generic direct and reverse reconciliation (DR and RR respectively) Gaussian CV QKD protocol has the form \cite{exp2,renner}:
\begin{equation}
\label{generic}
I_{DR}=\beta I_{AB}-\chi_{AE},\,\,\,I_{RR}=\beta I_{AB}-\chi_{BE}
\end{equation}
where $I_{AB}$ is the mutual information between trusted parties, Holevo quantity $\chi_{AE}$ ($\chi_{BE}$) is the upper bound on information, which can be available to an eavesdropper under the collective attacks scenario, and $\beta$ is the reconciliation efficiency, which characterizes, how effectively trusted parties process the data. The security of a scheme is established as the positivity of the lower bound (\ref{generic}) in either reconciliation scenario.

The successful experimental demonstrations of the CV QKD were performed \cite{exp2,exp1,exp3,exp4}, but two serious limitations were revealed. First, the DR scenario was remarkably shown to be unable to provide security under channel losses higher than $-3 dB$ \cite{coh2} (the limitation, which can be overcome by the use of either post-selection \cite{exp3} or the RR scheme \cite{coh3}). On the other hand, existing reconciliation algorithms suffer from low efficiency (the effect, also known in the discrete-variable QKD \cite{realrecon}), especially under RR and high level of loss \cite{exp2}. To overcome this limitation, the discrete modulation of coherent states \cite{discrete,discrete1,discrete2,discrete3,discrete4} can be used. Alternatively, it might be expected, that squeezing can qualitatively help when the reconciliation efficiency is limited and thus lowers $I_{AB}$. Since the analysis of DR and RR squeezed state protocol upon limited $\beta$ has not been carried out in \cite{sq1,sq2,sq3,sq4,sq5,coherent1} and squeezing was not distinguished from high modulation, it is not clear how much we gain from the feasible amount of squeezing.
 
In this paper, we predict that feasible limited squeezing is sufficient to reach evident tolerance to the channel noise for the long-distance quantum channels when the signal-to-noise ratio and, accordingly, reconciliation efficiency are low. To prove it, we have modified the entanglement-based scheme \cite{equiv} to be able to distinguish between the resources by independently varying squeezing and displacement in the state preparation. Thus we compare squeezed-state and coherent-state Gaussian protocols with optimal displacement under conditions of the same low reconciliation efficiency. We show the linear improvement of the security region by squeezing, which results in the fact that even limited squeezing can provide robustness to noise about an order of magnitude higher than is tolerable by coherent states. The result principally opens an experimentally feasible Gaussian-modulation alternative to the discrete modulation protocols, recently proposed for the same purpose of overcoming limited reconciliation.

\section{Modified entanglement-based scheme}

\begin{figure}
\centerline{\psfig{width=8.0cm,angle=0,file=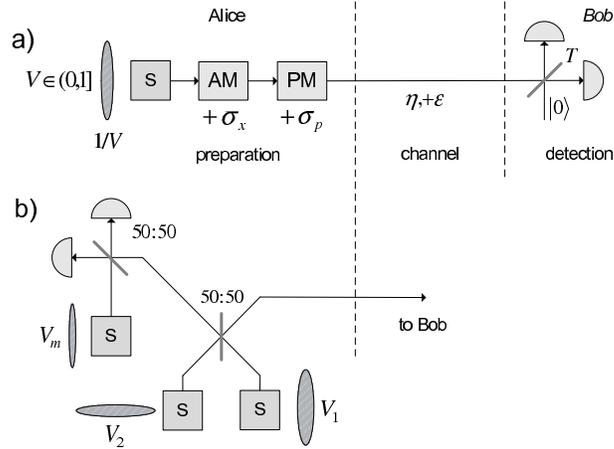}}
\caption{a) Continuous-variable quantum key distribution scheme: Alice generates squeezed signal states with squeezed variance $V$ and applies independent and generally different amplitude and phase quadrature displacement to encode each next bit using amplitude and phase quadrature modulators ($AM$ and $PM$). The signal travels through an untrusted lossy and noisy channel and is measured by Bob using two homodyne detectors, for $T=1$ the measurement is homodyne, and for $T=1/2$ it is heterodyne. b) Equivalent general source model based on the coupling of two oppositely-squeezed beams with different squeezing parameters and heterodyne measurement at the Alice side, fed by a squeezed state.}
\label{scheme_general}
\end{figure}

We verify the security of squeezed and coherent states in the conditions of an untrusted channel with transmittance $\eta$ and excess noise $\epsilon$. Since we estimate the maximum performance of a state in the Gaussian CV QKD, we assume that the state preparation is noiseless. At the same time we follow the pessimistic scenario and consider all the noise, observed by the remote party, as untrusted (it can be both channel and untrusted detector noise). 

We consider the displacement and measurement of quadrature observables of a light mode, which can be introduced as linear combination of anihilation and creation operators of the given mode as $x=a^\dag+a$ and $p=i(a^\dag-a)$. Under such a definition the quadrature fluctuations of a vacuum or coherent state mode will be equal to $1$, which we refer to as the shot-noise unit (SNU).

To access both displacement and squeezing independently in the conditions of noisy channel we perform calculations modeling the general prepare-and-measure scheme (see figure \ref{scheme_general} (a)) by the equivalent entanglement-based (EPR) scheme (figure \ref{scheme_general} (b)) representing a slightly modified version of the original scheme from \cite{equiv}. In this scheme the source is constructed by coupling two differently oppositely squeezed modes, while the measurement of Alice is carried out as heterodyne with the second input of the beam splitter fed by a squeezed vacuum with the squeezing direction being the same as measured by Alice. With varying the squeezing value, either coherent or squeezed state is conditionally prepared in Bob's mode. The correspondence between the modified EPR-based scheme and the generalized preparation scheme in terms of variances reads
\begin{equation}
V_{1,2}=V+\sigma_x\pm\sqrt{\frac{(V+\sigma_x)(\sigma_x+V\sigma_p(V+\sigma_x))}{1+V\sigma_p}}
\end{equation}
\begin{equation}
V_m=\frac{V^2\sigma_p(V+\sigma_x)}{\sigma_x(1+V\sigma_p)},
\end{equation}

where $\sigma_x$ and $\sigma_p$ are variances of, respectively, amplitude and phase quadrature displacement, and $V$ is squeezed variance in the prepare-and-measure scheme; $V_1$ is squeezed and $V_2$ is anti-squeezed variance of the states, used to construct the entangled state, and $V_m$ is the squeezed variance of the state put to the free port of a heterodyne detector at Alice side in the EPR-based scheme.

The mutual Shannon information between parties is the same for both DR and RR scenarios and is calculated using variances and conditional variances of the respective mode quadratures as $I_{AB}=1/2\log_2{V_B/V_{B|A}}=1/2\log_2{V_A/V_{A|B}}$, where $V_A=1/4(V_1+V_2+2V_m)$ is the variance of Alice's mode, $V_B=\eta/2(V_1+V_2+2\epsilon)+1-\eta$ is Bob's mode variance after the channel, and $V_{B|A}=V_B-C_{AB}^2/V_A$, where $C_{AB}=\sqrt{\eta}(V_2-V_1)/2\sqrt{2}$ is the correlation between Alice and Bob, so that

\begin{equation}
V_{B|A}=\frac{1}{2}\bigg(2-\frac{\eta(V_1-V_2)^2}{V_1+V_2+2V_m}+\eta(V_1+V_2+2\epsilon-2)\bigg)
\end{equation}

(similarly for $V_{A|B}$). Without loss of generality, we assume that $x$-quadrature is squeezed in the prepare-and-measure scenario and is, accodringly, measured by Alice in the entanglement-based scheme. In the case, when the squeezing direction is orthogonally rotated, the results for the squeezing and measurement of $p$-quadrature are directly accesible by transitions $V_1 \to 1/V_1$, $V_2 \to 1/V_2$ and $V_m \to 1/V_m$. 

In terms of the prepare-and-measure generalized scheme, the mutual information between the trusted parties reads:
\begin{equation}
\label{mi}
I_{AB}^x=\frac{1}{2}\log_2{\bigg[1+\frac{\sigma_x\eta}{\lambda+\eta(V+\epsilon-1)}\bigg]},
\end{equation}
where $\lambda=1$ corresponds to homodyne detection and $\lambda=2$ to heterodyne (note that obtaining classical information from the measurements of anti-squeezed quadrature is ineffective, unless the state is close to coherent). However, in the following exposition we consider only homodyne measurement. The reason for this is that heterodyne, being advantageous under RR and disadvantageous under DR, can be regarded as the particular case of "fighting noise with noise" \cite{sq4} when trusted noise at the Bob side decouples Eve from the measurement results. This effect is practically independent of squeezing (being valid also for coherent sates if bases are switched) and does not change the results below, which reveal the role of limited squeezing in CV QKD.

As was mentioned, Eve's information under the assumption of collective eavesdropping strategy is upper limited by the Holevo quantity $\chi_{AE}$ ($\chi_{BE}$ upon RR). The Holevo quantity can be written as $\chi_{AE}=S_E-\int P(A)S_{E|A}dA$, where $S_E$ is the von Neumann entropy of the eavesdropper state $\rho_E$ \cite{exp2}. The quantity $S_{E|A}$ is the von Neumann entropy of the eavesdropper state $\rho_{E|A}$ conditioned by the reference side measurement result $A$ and $P(A)$ is the distribution of the measured results. In case of Gaussian modulation of the coherent or squeezed states the entropy $S_{E|A}$ does not depend on Alice's measurement results and the Holevo quantity is expressed as $\chi_{AE}=S_E-S_{E|A}$ (similarly, $\chi_{BE}=S_E-S_{E|B}$ in the reverse reconciliation scenario) \cite{purification}. If channel noise is present, it is assumed that in the worst-case scenario the eavesdropper can purify the state shared between the trusted parties, thus $S_E=S_{AB}$, where $S_{AB}$ is the entropy of the state between Alice and Bob, and $S_{E|A}=S_{BC|A}$, and $S_{BC|A}$ is calculated from the conditional two-mode state (including Alice's heterodyne mode $C$) after Alice's projective measurement. Similar
considerations apply for the case of RR and give slightly  simpler expression $\chi_{BE}=S_{AB}-S_{A|B}$ (since Alice's heterodyne measurement does not change the entropies and taking mode $C$ into account is not necessary), with the latter term being calculated from the state shared between Alice and Bob conditioned on Bob's measurement.

The covariance matrix $\gamma_{AB}$ of the two-mode Alice-Bob state after heterodyne at Alice is given by:

\begin{equation}
\gamma_{AB}=
\left( \begin{array}{cccc}
V_A^x & 0 & \sqrt{\eta}C_{AB}^x & 0 \\
0 & V_A^p & 0 & \sqrt{\eta}C_{AB}^p \\
\sqrt{\eta}C_{AB}^x & 0 & \eta(V_B^x+\epsilon-1)+1 & 0 \\
0 & \sqrt{\eta}C_{AB}^p & 0 & \eta(V_B^p+\epsilon-1)+1
\end{array} \right),
\end{equation}

where source variances are $V_A^x=(V_1+V_2+2V_m)/4$, $V_A^p=1/(4V_1)+1/(4V_2)+1/(2V_m)$, $V_B^x=(V_1+V_2)/2$, $V_B^p=1/(2V_1)+1/(2V_2)$, and source correlations are $C_{AB}^x=(V_2-V_1)/(2\sqrt{2})$ and $C_{AB}^p=(1/V_2-1/V_1)/(2\sqrt{2})$.

The conditional matrix $\gamma_A^{x_B}$ after Bob's homodyne measurement in $x$, which is used in the calculation of the RR scenario, is obtained (\cite{exp2}) as

\begin{equation}
\label{condmat}
\gamma_A^{x_B}=\gamma_A-\sigma_{AB}(X \gamma_B X)^{MP}\sigma_{AB}^T,
\end{equation}

where $\gamma_A$ , $\gamma_B$ and $\sigma_{AB}$ are the submatrices of $\gamma_{AB}$, standing for individual modes and their correlation, $MP$ denotes the Moore-Penrose pseudoinverse of a matrix and $X$ is the $2\times 2$ matrix, with elements equal to $0$, except the first, which is $1$. The resulting matrix $\gamma_A^{x_B}$ is given by

\begin{equation}
\gamma_A^{x_B}=
\left( \begin{array}{cc}
V_A^x-\frac{\eta(V_1-V_2)^2}{8+4\eta(V_1+V_2+2\epsilon-2)} & 0 \\
0 & V_A^p
\end{array} \right).
\end{equation}

The entropies then are given by the expressions $S_{AB}=G[(\lambda_1-1)/2]+G[(\lambda_2-1)/2]$ and $S_{A|B}=G[(\lambda_3-1)/2]$, where $\lambda_{1,2}$ are the symplectic eigenvalues of the matrix $\gamma_{AB}$, $\lambda_3$ is the symplectic eigenvalue of $\gamma_A^{x_B}$, and $G(x)=(x+1)\log (x+1)-x\log x$. Similar calculations apply to the DR scenario, when $S_{BC|A}$ is calculated from two symplectic eigenvalues of the covariance matrix $\gamma_{BC}^{x_A}$ after Alice's measurement in $x$.

In the simplest case of a purely lossy channel the key rate can be relatively easy calculated analytically. Indeed, the channel in this case can be described as a beam splitter with transmittance $\eta$. Then the state, which is available to Eve for collective measurement, is the state of the second output mode of the beam splitter. The von Neumann entropies, contributing to the Holevo quantity, are directly calculated from two symplectic eigenvalues $\lambda_{1,2}$ of the two $2 \times 2$ covariance matrices, describing Eve's mode state $\rho_E$ before and $\rho_{E|A}$ (or $\rho_{E|B}$ depending on reconciliation direction) after the respective trusted party measurement (calculated similarly to (\ref{condmat})), resulting in $\chi=G[(\lambda_1-1)/2]-G[(\lambda_2-1)/2]$. The symplectic eigenvalues are then the square roots of covariance matrix determinants, which, since matrices are diagonal, are the products of matrix elements, and, in the more relevant case of reverse reconciliation, read: 

\begin{eqnarray}
\label{lambda1pureloss}
\lambda_1=\prod_{i=\{x,p\}}{\sqrt{V_B^i(1-\eta)+\eta}} \\
\label{lambda2pureloss}
\lambda_2=\sqrt{\Bigg(V_B^x(1-\eta)+\eta-\frac{(C_{BE}^x)^2}{V_B^x\eta+1-\eta}\Bigg)\bigg(V_B^p(1-\eta)+\eta\bigg)},
\end{eqnarray}

where $C_{BE}^{x,p}=\sqrt{\eta(1-\eta)}(1-V_B^{x,p})$ is the correlation between Bob's and Eve's modes in $x(p)$ quadrature (note, that similar calculations apply to the measurement of $p$ quadrature). Taking into account the expressions for mode variances, after straightforward calculations these symplectic eigenvalues directly give expression for the Holevo quantity, and, being combined with mutual information (\ref{mi}) lead to the analytical expression for the lower bound on the key rate (\ref{generic}) under pure channel loss.

In the presence of channel noise the calculations were performed semianalytically with numerical estimations in the general case and analytical expressions in the limiting cases. Further we estimate the key rate and establish the security bounds in terms of the tolerable channel noise, which is the main objective limitation for security of the protocols, since any loss can be tolerated upon the RR.

\section{Squeezed state quantum key distribution (QKD) with limited reconciliation efficiency}

While in \cite{sq1,sq2,sq3,sq4,sq5,coherent1} the reconciliation efficiency is perfect, we focus here on limited reconciliation efficiency. This must be taken into account, since it turns out to be one of the main practical limiting factors for the Gaussian CV QKD, especially if RR is used. In particular, the efficiency $\beta \in (0.8,0.9)$ was reached in the recent experimental implementation of the Gaussian CV QKD with coherent states and RR \cite{exp2}, and thus the scheme was limited by the 25 km distance (corresponding to transmittance $\eta \approx 0.3$), while displacement had to be optimized. 

\begin{figure}[h]
\begin{tabular}{ll}
\includegraphics[width=0.4\textwidth]{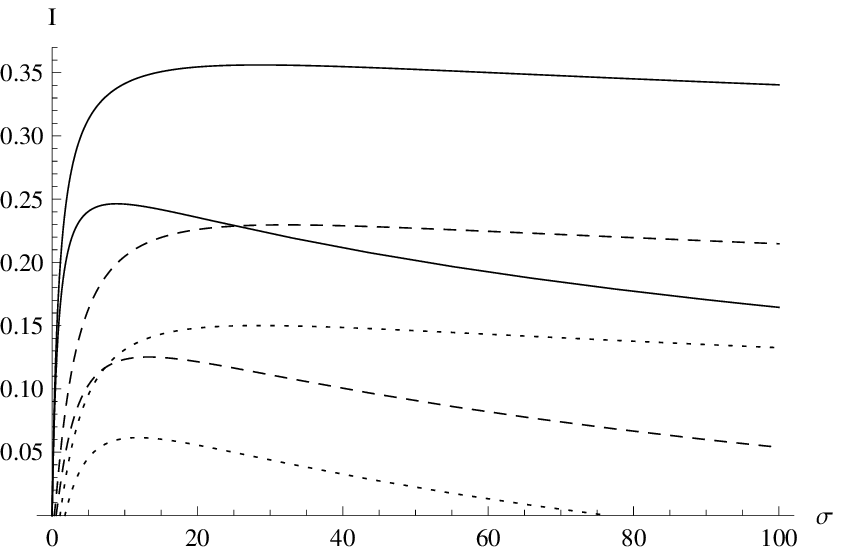}
\includegraphics[width=0.4\textwidth]{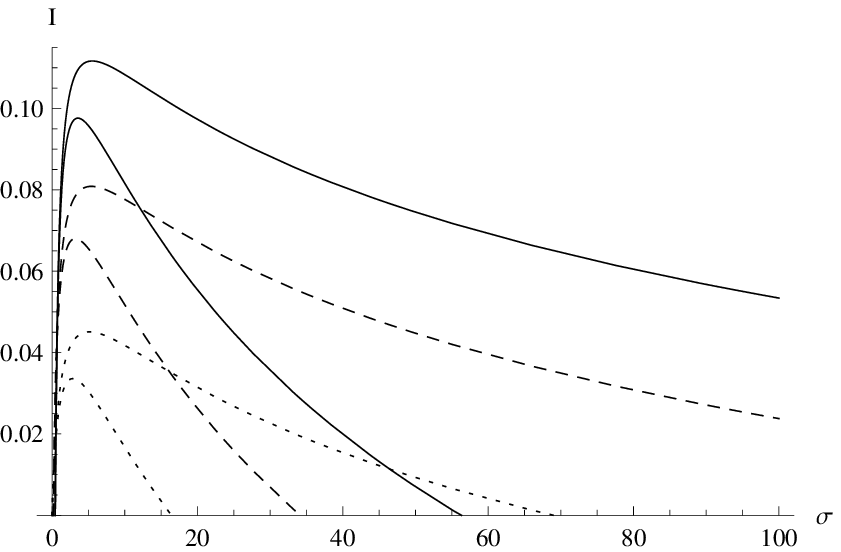}
\end{tabular}
\caption{Lower bound on the key rate secure against collective attacks in the DR (left) and RR (right) scenarios for coherent state $V=1$ (dotted lines), moderately squeezed states $V=0.5$ (dashed lines) and strongly squeezed states $V=0.1$ (solid lines). Upper lines correspond to reconciliation efficiency $\beta=0.95$, lower lines correspond to $\beta=0.9$. Channel transmittance is $\eta=0.6$ for DR and $\eta=0.1$ for RR.
\label{KRvsSigma}}
\end{figure}

Although reconciliation efficiency can be relatively high in the case of DR, we start from this scenario. First, on figure \ref{KRvsSigma} (left) we show how the secure key rate depends on displacement variance for different values of squeezing and reconciliation efficiency. It is evident from the plots that even moderate squeezing ($-3 dB$) quantitatively improves the key rate, while displacement variance must be optimized. 

\begin{figure}[h]
\begin{tabular}{ll}
\includegraphics[width=0.4\textwidth]{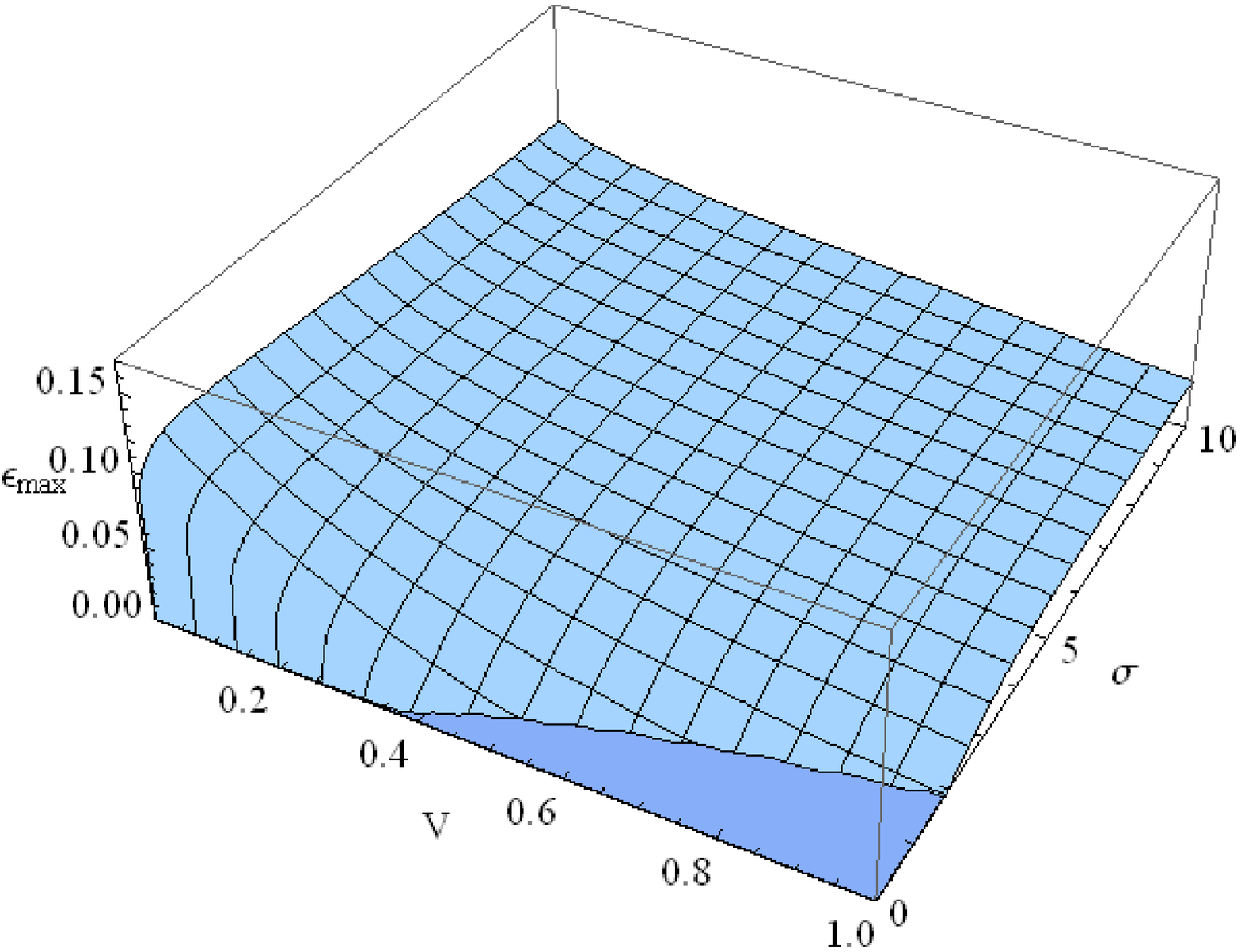}
\includegraphics[width=0.4\textwidth]{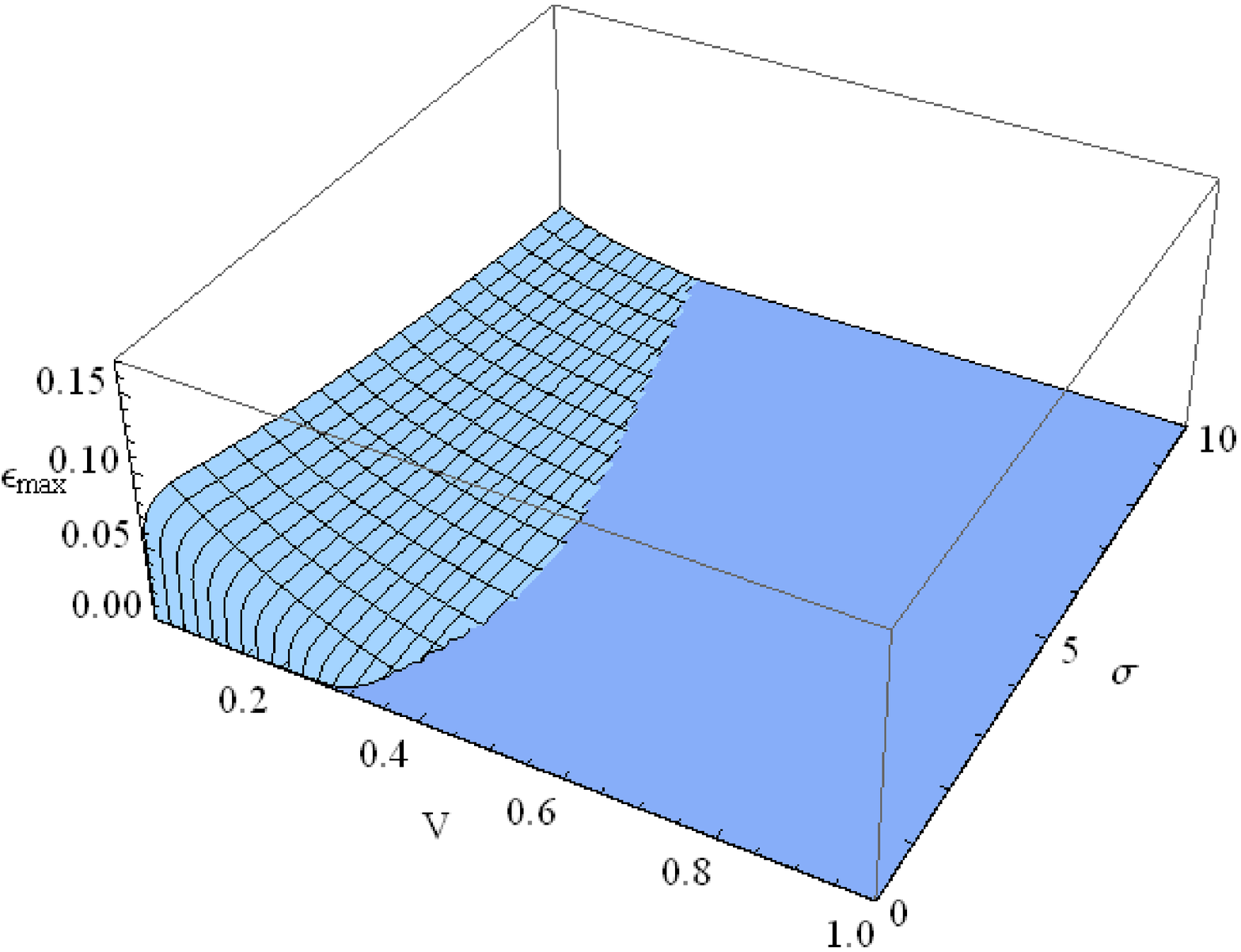}
\end{tabular}
\begin{tabular}{ll}
\includegraphics[width=0.4\textwidth]{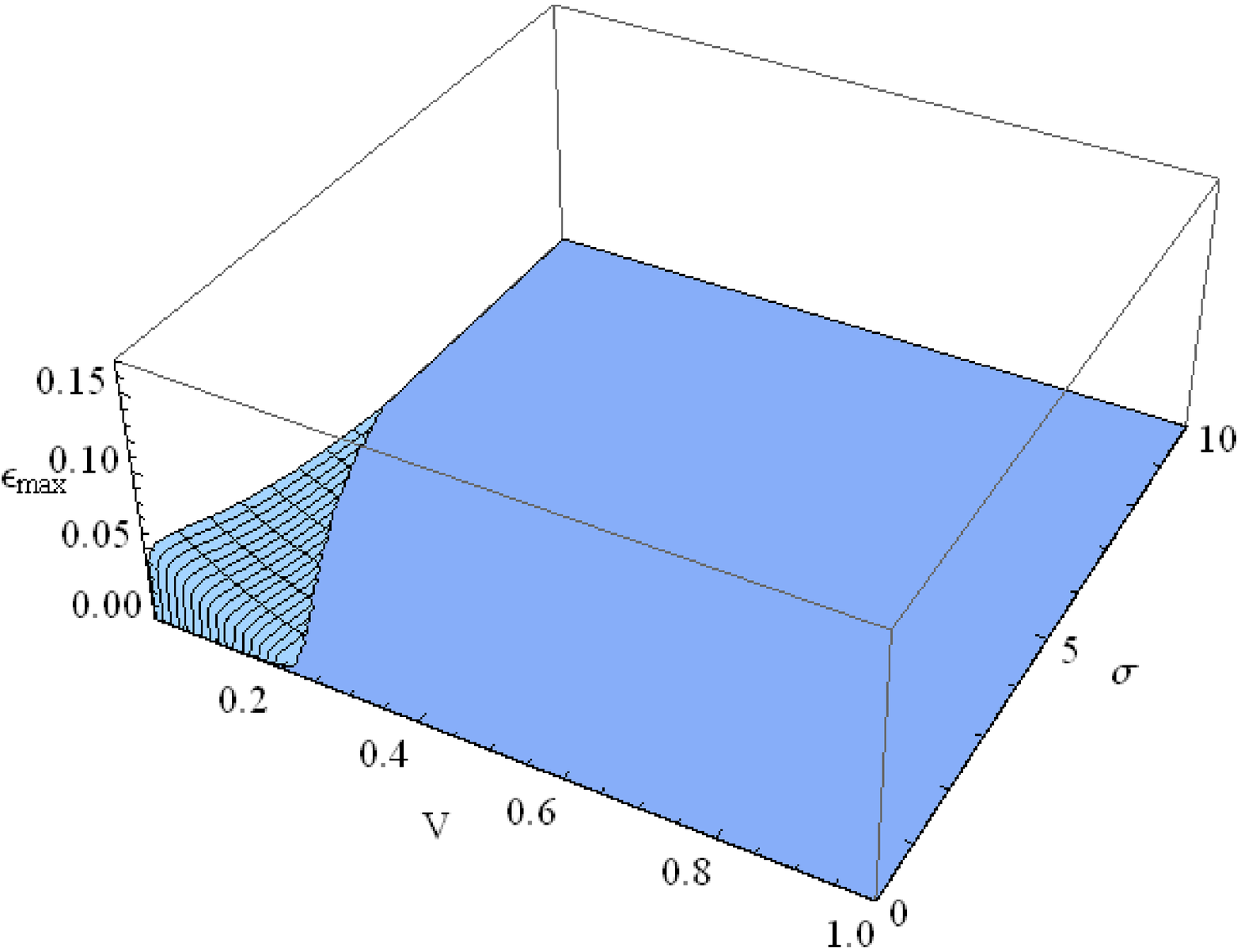}
\includegraphics[width=0.4\textwidth]{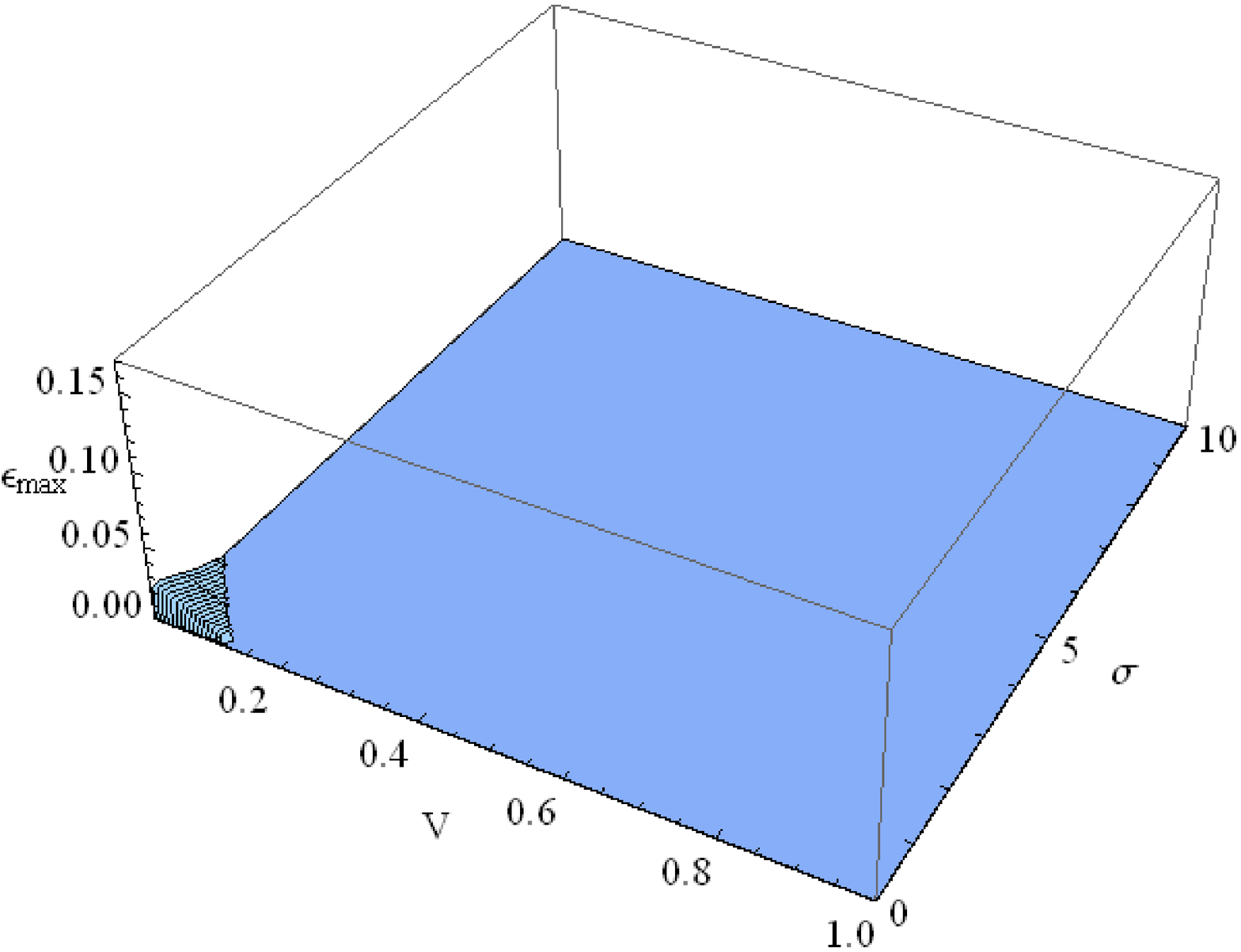}
\end{tabular}
\caption{Security region for the CV QKD scheme upon DR versus squeezed variance $V$ and displacement variance $\sigma$ for channel transmittance $\eta=0.6$ and reconciliation efficiency $\beta=0.9,0.8,0.7,0.6$ (from left to right, from top to bottom).
\label{efficDR}}
\end{figure}

Similar behaviour is demonstrated by the security region in terms of the tolerable channel noise, which also depends on squeezing and displacement (see figure \ref{efficDR}).

\begin{figure}[h]
\begin{tabular}{ll}
\includegraphics[width=0.4\textwidth]{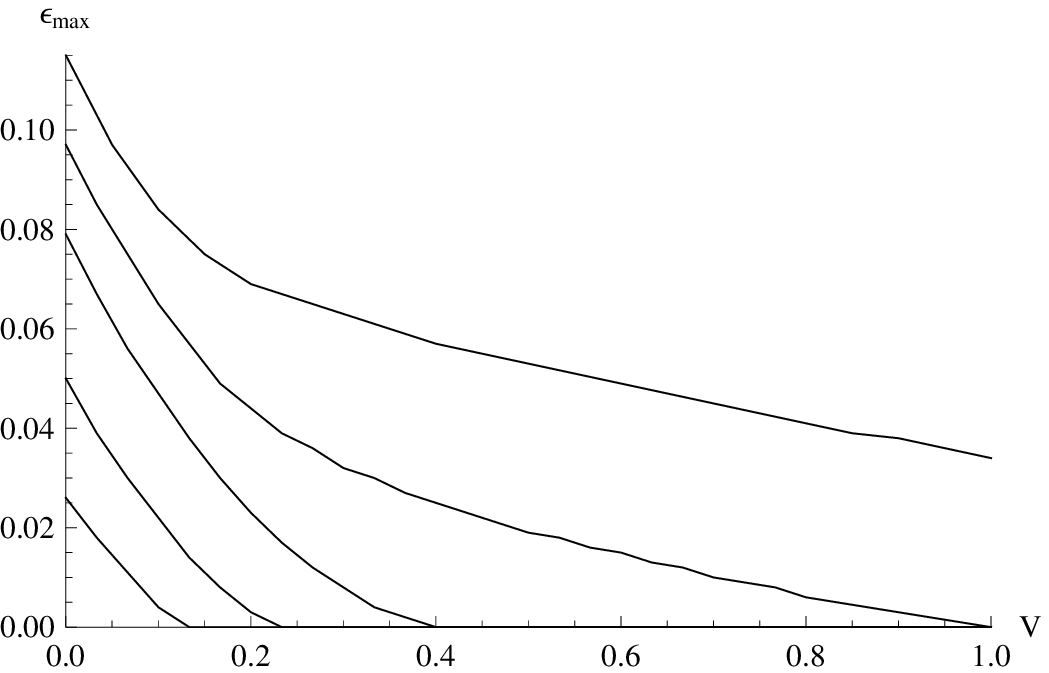}
\includegraphics[width=0.4\textwidth]{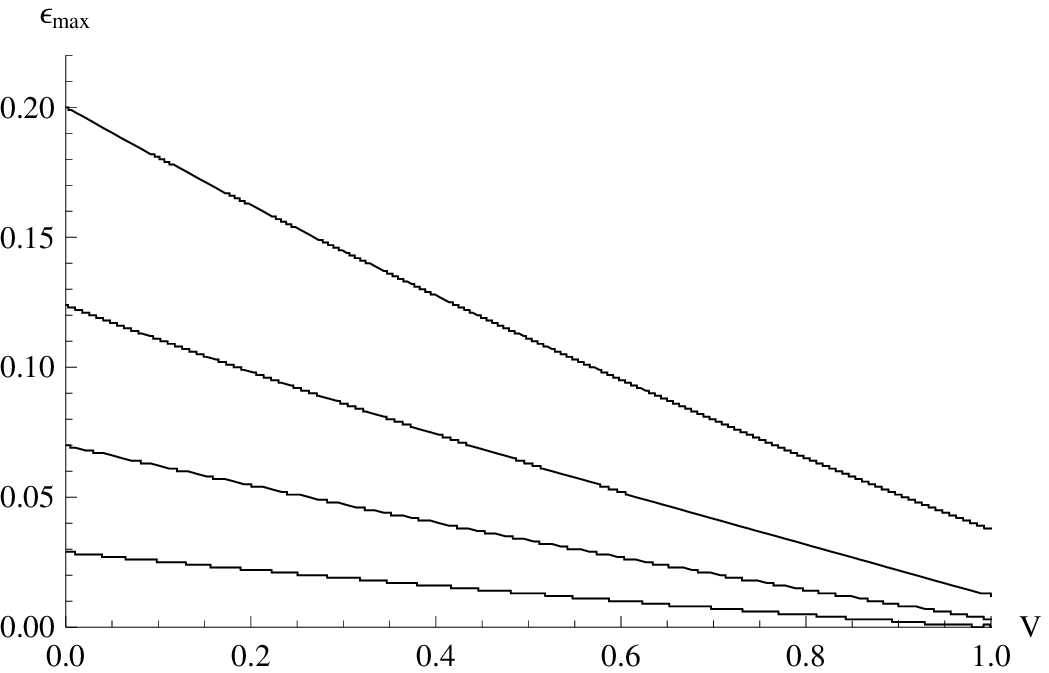}
\end{tabular}
\caption{Maximum tolerable channel excess noise versus squeezing $V$ upon optimized displacement variance $\sigma$, applied to both quadratures. Left: DR, reconciliation efficiency $\beta=0.9,0.85,0.8,0.7,0.6$ (from top to bottom), channel transmittance $\eta=0.6$; right: RR,  reconciliation efficiency $\beta=0.8,0.6,0.4,0.2$ (from top to bottom), channel transmittance $\eta=0.1$.
\label{emax_vs_v_opt}}
\end{figure}

As was already analyzed in \cite{sq5}, upon perfect reconciliation, limited squeezing with low modulation (small displacement variance) improves the tolerance of the DR scheme to loss, while infinitely strong squeezing is disadvantageous from the point of view of tolerable channel noise, compared to the coherent states. In contrast to this, upon imperfect reconciliation, squeezing continuously improves the security region of the protocol, while displacement must be limited and optimized to provide maximum tolerance to channel noise for given signal state variance. Optimization was performed for the case of DR at fixed values of $\beta$ with the results given in figure \ref{emax_vs_v_opt} for low channel transmittance. The results evidently demonstrate the positive role of finite squeezing for DR upon imperfect reconciliation.

\begin{figure}
\centerline{\psfig{width=8.0cm,angle=0,file=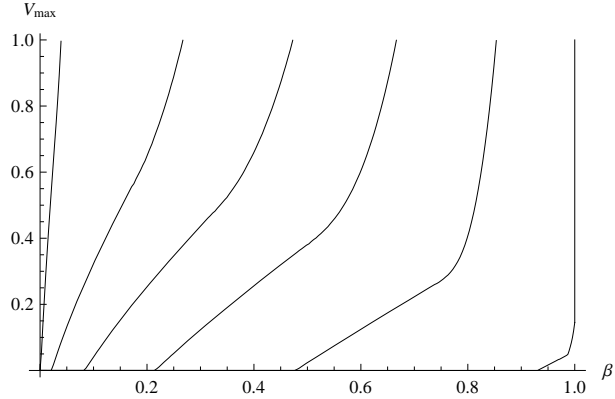}}
\caption{Maximum squeezed variance of the signal states, providing a positive key rate, upon direct reconciliation and optimized displacement, for the given reconciliation efficiency $\beta$ and different values of channel transmittance $\eta=0.99,0.9,0.8,0.7,0.6,0.51$ (from left to right).}
\label{vmax_vs_b_opt-DR}
\end{figure}

Moreover, when efficiency $\beta$ is low enough, squeezing appears to be strictly required to establish secure quantum communication, as Gaussian modulated coherent states fail to provide a positive key rate even upon purely lossy channel. This restriction was analytically obtained already for the case of individual attacks on coherent states-based protocol when security can be provided only for reconciliation efficiency within the security constraint $\beta > 1/\eta - 1$, $\eta \in (0.5,1)$ (for optimized displacement, giving the highest key rate). For the case of collective attacks, the maximum variance of the signal states, providing positive key rate, upon optimized displacement, was calculated numerically and is given in figure \ref{vmax_vs_b_opt-DR}. It is clearly seen, that squeezing of the states becomes required upon reduced reconciliation efficiency and low channel transmittance, and the necessity in squeezing is enforced by a decrease of both of the parameters. Thus, in the respective region of parameters, Gaussian DR CV QKD protocol enters strictly the nonclassical regime when only squeezed states can provide security of the key, even under pure channel loss.

However, since the DR scenario is more effective in terms of reconciliation and it cannot tolerate transmittances lower than $0.5$, it is more important to analyze how limited reconciliation affects the RR scheme. Indeed, the RR protocols, already being less effective in data processing, are aimed to perform upon highly lossy long-distance channels, while losses additionally decrease signal-to-noise ratio (SNR) and, accordingly, $\beta$. The explicit dependence of $\beta$ on SNR is the intrinsic feature of a particular scheme, as it also depends on the effectiveness of classical algorithms used for Gaussian data binarization and subsequent error correction. The practical reconciliation codes are known to provide less than $50\%$ efficiency for SNR around 1 \cite{recon}. However, in practice (from the implementation of a coherent protocol \cite{exp2}) it is known that the key rate is the convex function of SNR for $\beta<1$ and the lower the reconciliation efficiency, the lower the optimal displacement variance. Moreover, for reduced efficiency the displacement must be limited. For $\beta=0.7$, for example, the maximum applicable displacement variance for coherent states is approximately 4 SNU, which, in turn, makes SNR about $0.4$ after the channel with $10\%$ transmittance, and this additionally lowers the achievable reconciliation efficiency below $50\%$. Thus, it is generally reasonable to consider low values of $\beta$ in order to clarify the role of the resources in the applicability of Gaussian QKD under RR scenario. 

\begin{figure}[h]
\begin{tabular}{ll}
\includegraphics[width=0.4\textwidth]{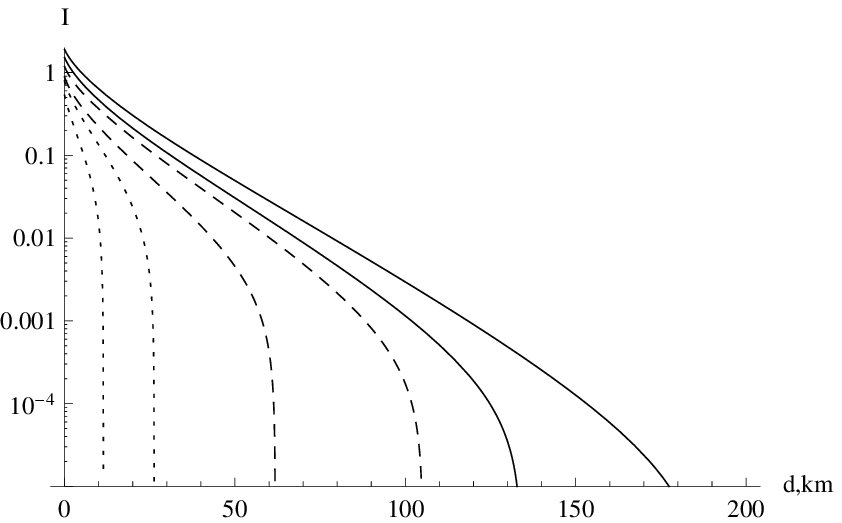}
\includegraphics[width=0.4\textwidth]{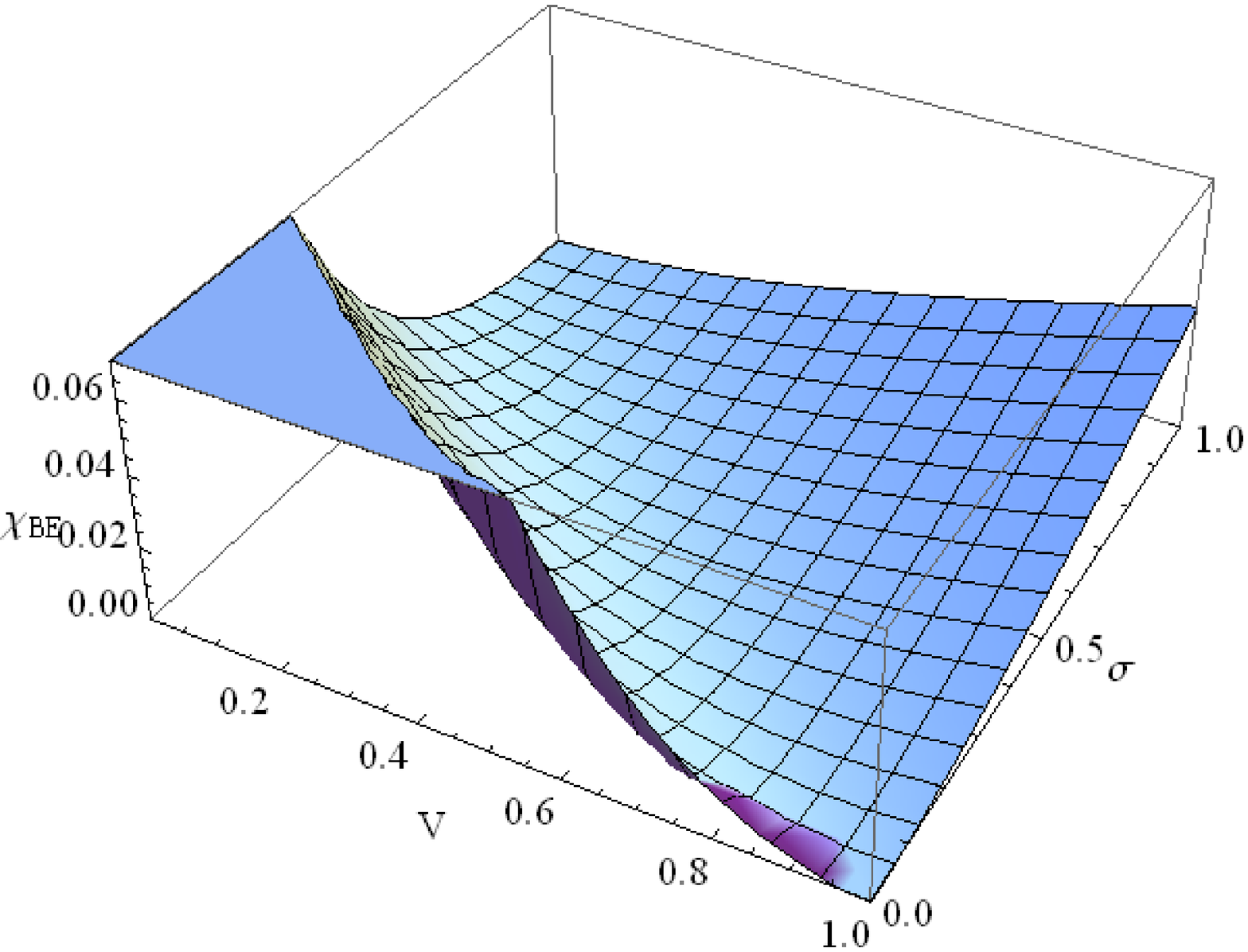}
\end{tabular}
\caption{Left: lower bound on the key rate secure against collective attacks in the RR scenario versus distance (in fiber with standard attenuation of -0.2 dB/km) for coherent $V=1$ (dotted lines), moderately squeezed $V=0.5$ (dashed lines) and strongly squeezed $V=0.1$ (solid lines) signal states upon optimized displacement variance. Upper lines correspond to reconciliation efficiency $\beta=0.95$ and lower lines correspond to $\beta=0.9$. Channel noise is fixed at $\epsilon=0.1$. Right: profile of the Holevo quantity, upper limiting the information, available to Eve in the RR scheme versus state squeezing $V$ and applied displacement variance $\sigma$ for low channel transmittance $\eta=0.1$ and no channel noise. 
\label{KRRRandHolevo}}
\end{figure}

The results of the numerical estimation of the key rate for the reduced reconciliation efficiency of the reverse protocol under purely lossy channel with transmittance $\eta=0.1$ are given in figure \ref{KRvsSigma} (right). Similarly to the case of DR, squeezing quantitatively improves the key rate already for the pure channel loss, and under reduced efficiency displacement variance must be optimized to provide maximum key rate. In case when displacement is optimized, even moderate squeezing is shown to improve the key rate values and the maximum distance of the Gaussian protocol upon the given fixed channel noise $\epsilon$, as seen from figure \ref{KRRRandHolevo} (left). 

\begin{figure}[h]
\begin{tabular}{ll}
\includegraphics[width=0.4\textwidth]{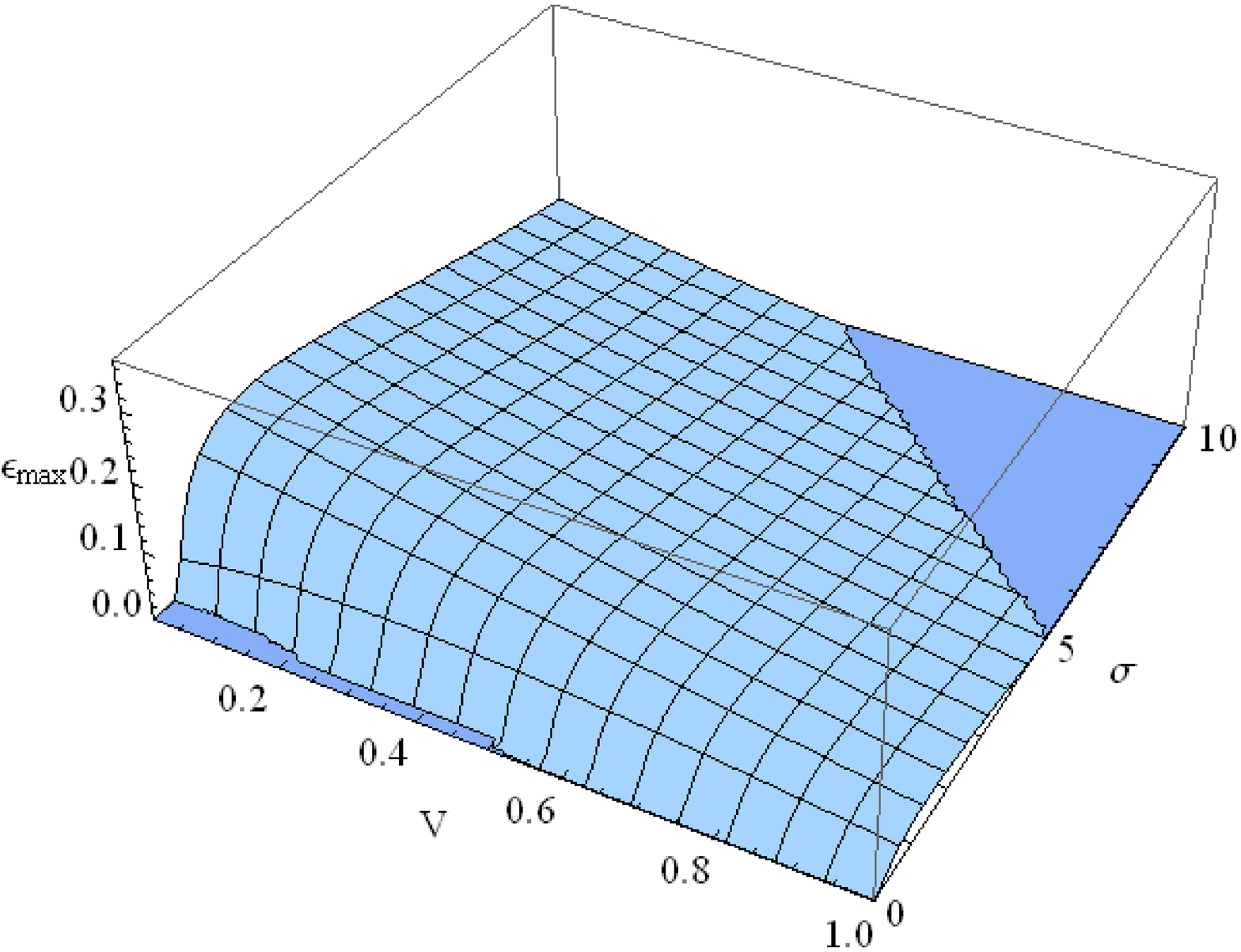}
\includegraphics[width=0.4\textwidth]{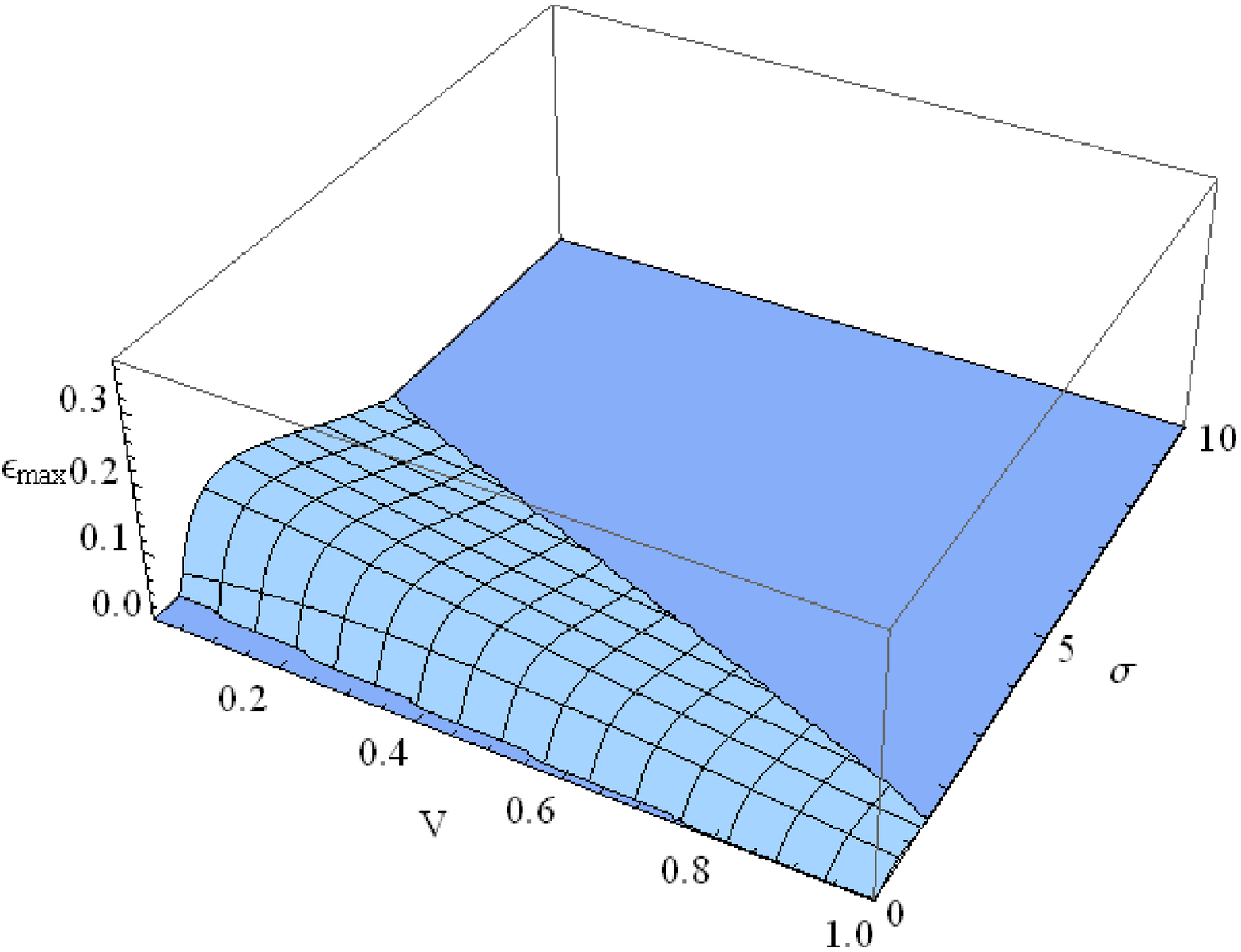}
\end{tabular}
\begin{tabular}{ll}
\includegraphics[width=0.4\textwidth]{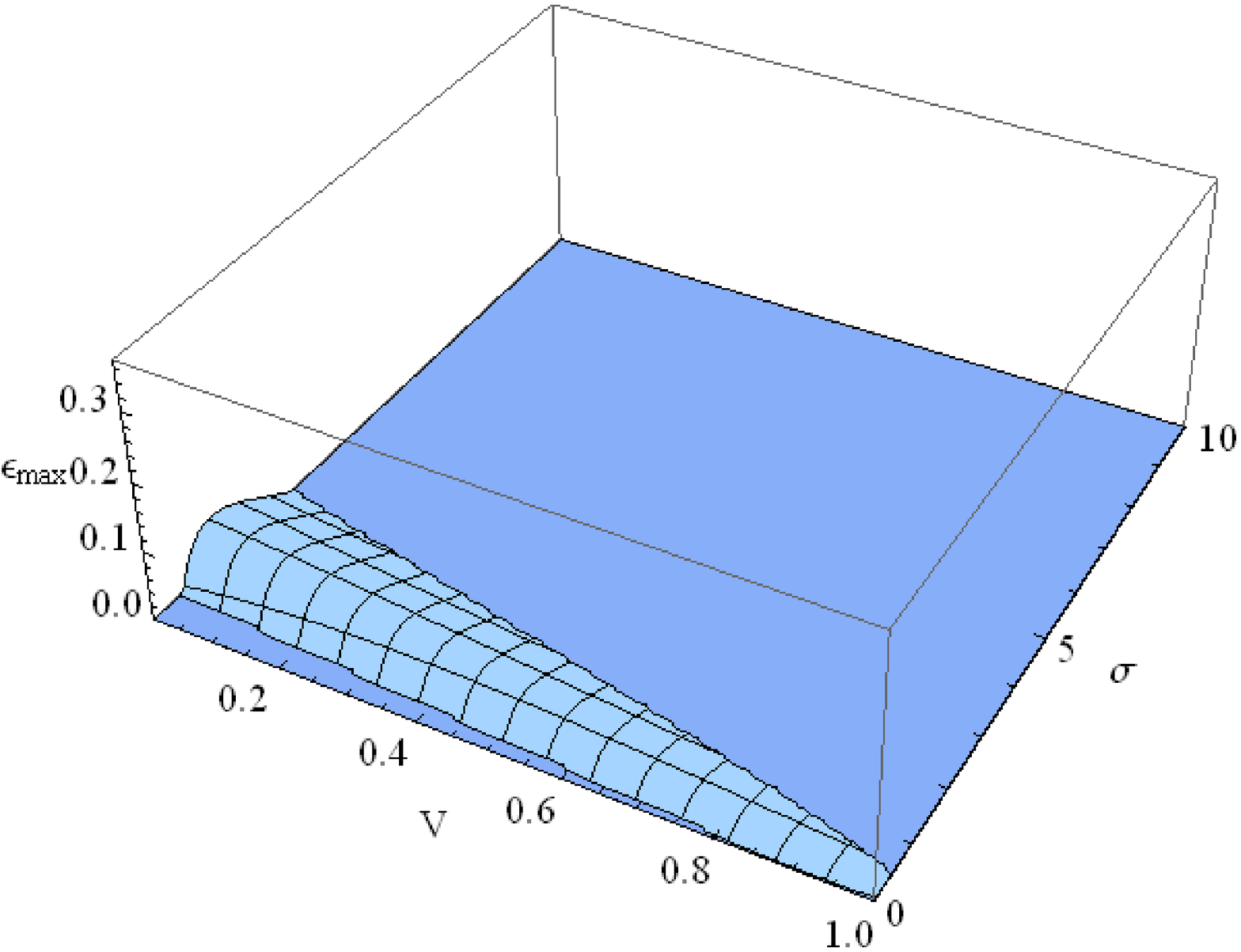}
\includegraphics[width=0.4\textwidth]{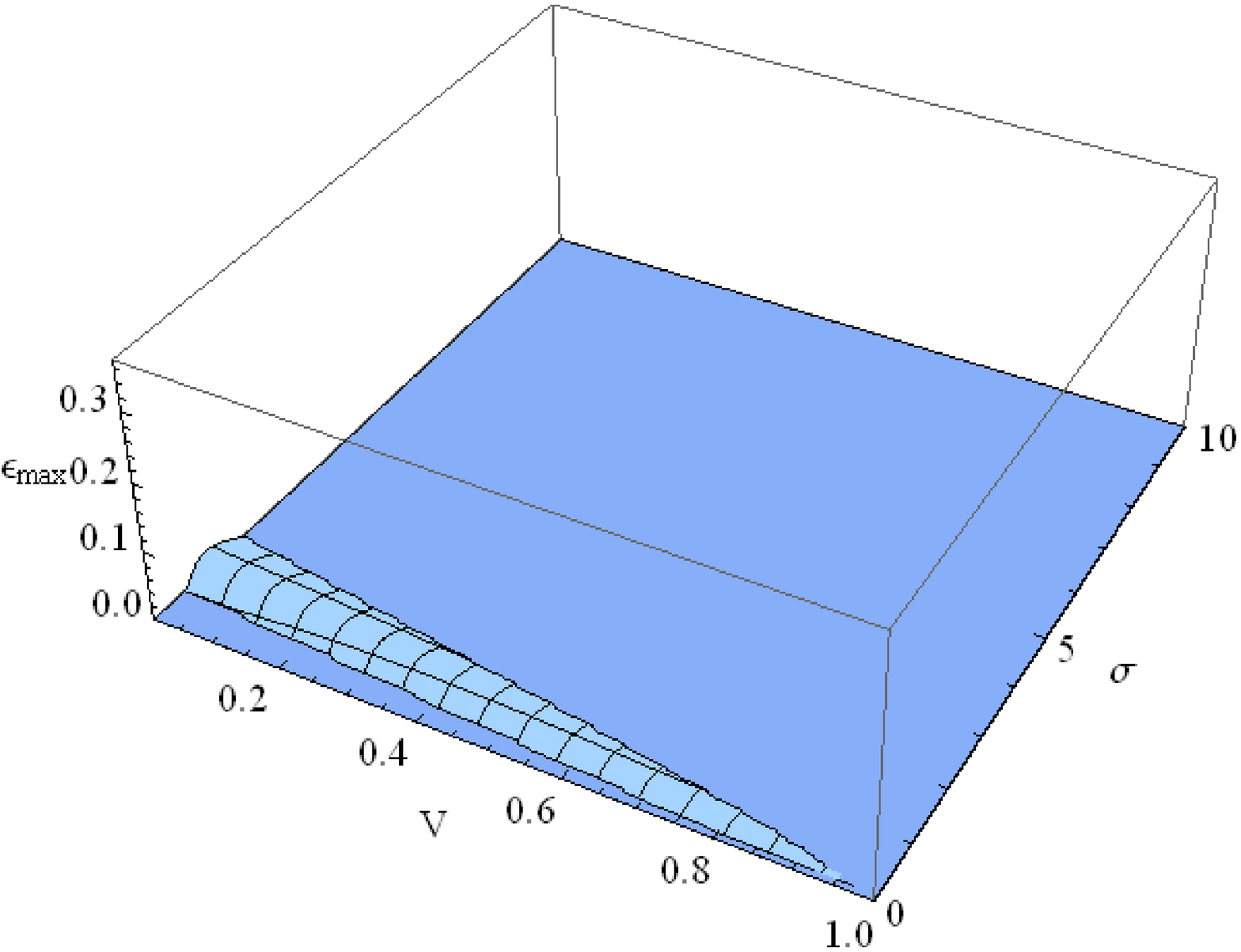}
\end{tabular}
\caption{Security region for the CV QKD scheme upon RR versus squeezed variance $V$ and displacement variance $\sigma$ for channel transmittance $\eta=0.1$ and reconciliation efficiency $\beta=0.8,0.6,0.4,0.2$ (from left to right, from top to bottom).
\label{efficRR}}
\end{figure}

The security region of the RR Gaussian protocol in terms of the maximum tolerable channel noise demonstrates the analogous behaviour as seen from figure \ref{efficRR}: under reduced efficiency the security region is shrinking and displacement variance must be reduced as well (which is already known for a coherent protocol \cite{exp2}). Note, that "traditional" squeezed-based schemes \cite{sq1,sq2,sq3,sq4,sq5} with displacement variance $\sigma=1/V-V$ start to be inapplicable under limited $\beta$. However, it turns out that squeezed states allow higher displacement variance and provide higher tolerable excess noise. This is clearly visible from the profiles of the maximum tolerable channel excess noise under optimized displacement variance, given in figure \ref{emax_vs_v_opt} (right), for various values of $\beta$. While for relatively high $\beta=0.8$ and even $\beta=0.6$ coherent states ($V=1$) still show a reasonable offset in terms of the tolerable channel noise, for lower reconciliation efficiency it is negligibly small, at the same time even moderate squeezing provides tolerance to noise. Remarkably, the dependence on squeezing remains close to linear, i.e. squeezing is linearly improving maximum tolerable channel noise in the whole region of reconciliation efficiency. At the same time, under low efficiencies Gaussian modulated coherent states fail to demonstrate any noticeable tolerance to noise. 

In the limiting case of highly squeezed states $V \to 0$ the displacement variance must be in the range:

\begin{equation}
\frac{1}{1+\sqrt{\beta}} < \sigma \big|_{V\to 0} < \frac{1}{1-\sqrt{\beta}}
\label{sqlimits}
\end{equation}

The typical values of maximum tolerable channel excess noise $\epsilon$ upon fixed limited reconciliation efficiency $\beta$ and optimal displacement variance $\sigma$ for coherent states $V=1$ and moderately squeezed states $V=0.5$ ({\it i.e.}, -3 dB of squeezing, which is completely feasible with current technology, as -10 dB was already obtained \cite{sqexp}) are given in the table \ref{epstable}. It is evident from the table that at low reconciliation efficiency the limited squeezing can already achieve tolerance to the channel noise, which is about an order of magnitude higher than is tolerable by coherent states upon the same efficiency $\beta$ and the same channel transmittance. 

Thus, in the region of low efficiency $\beta$ and in the presence of evident untrusted noise, the Gaussian CV QKD protocol enters the nonclassical regime when at least moderate squeezing is required to enable secure key distribution.

\begin{table}
\caption{\label{epstable}Values of the maximum tolerable channel excess noise $\epsilon$ upon limited reconciliation efficiency $\beta$ and optimal displacement variance $\sigma$ for coherent states $V=1$ and moderately squeezed states $V=0.5$, channel transmittance is $\eta=0.1$}
\begin{indented}
\item[]\begin{tabular}{@{}lll}
\br
&\centre{2}{Maximum tolerable $\epsilon$}\\
\ns
&\crule{2}\\
\centre{1}{$\beta$}&$V=1$&$V=0.5$\\
\mr
$20\%$ &$10^{-3}$&$1.3\cdot 10^{-2}$\\
$40\%$ &$3.7\cdot 10^{-3}$&$3.4\cdot 10^{-2}$\\
$60\%$ &$1.2\cdot 10^{-2}$&$6.3 \cdot 10^{-2}$\\
$80\%$ &$3.8\cdot 10^{-2}$&$1.1\cdot 10^{-1}$\\
\br
\end{tabular}
\end{indented}
\end{table}

The reason for this effect is in the behavior of the Holevo quantity $\chi_{EB}$, describing the upper bound on the quantum information, available to an eavesdropper. Indeed, for low $\beta$ the contribution of mutual information to the lower bound (\ref{generic}) becomes very low; in addition, for low channel transmittances $\eta \ll 1$ the mutual information practically does not depend on squeezing as $I_{AB}=\sigma\eta/\log{4} + O[\eta]^2$. Thus, the question is: which kind of states is more effectively minimizing the Holevo quantity? Although the Holevo quantity is the entropic measure of maximum information, available to an eavesdropper, its minimization has direct physicsl meaning. 

For the purely lossy channels the Holevo quantity is minimized with the value $0$, which is reached when the two parts of the quantity $\chi_{BE}=S_E-S_{E|B}$ (calculated from Eve's state and Eve's state conditioned by Bob's measurement) are equal. From expressions (\ref{lambda1pureloss}) and (\ref{lambda2pureloss}) for symplectic eigenvalues, which are used to calculate both parts $S_E$ and $S_{E|B}$ of the Holevo quantity, it is evident that the eigenvalues (and, accordingly, the two parts of the Holevo quantity) are equal when correlation between Bob and Eve is equal to $0$. It is achieved, regardless of channel transmittance, when $V_B^x=1$, i.e. $\sigma_x=1-V$, where $V$ is the squeezed state variance in the prepare-and-measure scheme. The typical profile of $\chi_{EB}$ is given at figure \ref{KRRRandHolevo} (right) versus signal state squeezing and displacement variance upon purely lossy channel. 

In case when channel noise is present, the Holevo quantity is minimized upon the same value of modulation (the minimum value being positive, i.e. the typical plot of the Holevo quantity, given in figure \ref{KRRRandHolevo}, is moved up by the channel noise equivalently for any squeezing and modulation). It was confirmed numerically using the purification method, but can also be obtained from analytical considerations if entangling cloner attack is considered. Such an attack was shown to be the optimal individual attack \cite{equiv} and was recently used to calculate security against collective attacks \cite{weedbrook}. Under entangling cloner attack the correlation between Bob and Eve's cloner mode is again directly accessible, it is $\sqrt{\eta(1-\eta)}(1-V_B^x+\eta\epsilon)$, where $\epsilon$ is the channel noise. Again, this correlation is minimized upon $V_B^x$ being equal to $1$. 

Thus, the minimum of the Holevo quantity, i.e. the minimization of the upper bound on information, available to Eve, which is relevant in the low $\beta$ regime upon RR, is achieved upon complete (for pure channel loss) or maximum possible decoupling of Eve from Bob, when the correlation between their modes is minimized. It depends solely on the state properties before channel and is described by the simple condition $\sigma_x=1-V$. Evidently, squeezed states reach this de-correlation upon positive displacement variance, while coherent states minimize the Holevo quantity only when no displacement is applied. Moreover, in case when the reconciliation efficiency is arbitrarily small (i.e. $\beta \ll 1$) the security can be achieved only for $\sigma \leq 1-V$, which is evidently 0 for Gaussian modulated coherent states, i.e. no information can be added to coherent states by displacement, when reconciliation efficiency $\beta$ is low. At the same time even small squeezing of quantum states allows additional displacement and is thus able to carry information, contributing to the secure key.

Thereby, the case of low reconciliation efficiency represents the essentially nonclassical region of QKD, where squeezing plays the main role, whereas the contribution from the classical resources is minor. From this follows that nonclassical resource, namely squeezing, can partly substitute the limited classical reconciliation resource.



\section{Conclusions}

We have analyzed the role of squeezing in the continuous-variable quantum key distribution, clearly distinguishing squeezing from displacement in a generalized preparation scheme, taking into account imperfect reconciliation. We have shown that squeezing improves tolerance to untrusted noise in both direct and reverse reconciliation schemes and is strictly required in the direct case when reconciliation efficiency is low. For highly attenuating channels, corresponding to long-distance links, squeezing demonstrates linear improvement of the security region in terms of tolerable untrusted noise under optimized displacement. Hereby, even limitedly squeezed states can provide security in the conditions of a strongly attenuating channel and in the presence of untrusted noise, contrary to coherent states, which fail to demonstrate any non-negligible tolerance to noise under the same low reconciliation efficiency. The reason for the effect is the behavior of the Holevo quantity, upper bounding the eavesdropper's information. It turns out to be more effectively minimized by the squeezed states, i.e. under higher displacement, applied to the states.

Thus, under low reconciliation efficiency, both direct and reverse Gaussian protocols enter essentially the nonclassical regime, when tolerance to the noticeable channel noise can be provided only by squeezed states.  Weak Gaussian modulation of moderately squeezed states in the reverse scenario can be a feasible alternative to the discrete modulation of coherent states for the long-distance links. However, the most effective solution will probably be the combination of the two approaches with optimization of both modulation scheme and quantum state as the resource.

\ack
VCU was supported by the project no. P205/10/P321 of the Grant Agency of Czech Republic, RF was supported by the projects  no. MSM 6198959213, no. LC 06007 and no. ME 10156 of the Czech Ministry of Education, project no. 202/07/J040 of the Grant Agency of Czech Republic and the EU COMPAS project.

\end{document}